\documentclass[plb,twocolumn,nofootinbib]{revtex4-1}
\usepackage{graphicx}
\usepackage{cancel}
\usepackage{amssymb}
\usepackage{textcomp}
\usepackage{amsmath}
\usepackage{bm}
\usepackage{times}
\usepackage{epsfig}
\usepackage{color}
\usepackage{graphics}
\usepackage{hyperref}
\usepackage{setspace}
\usepackage{paralist}
\usepackage{physics}
\usepackage{slashed}
\usepackage{comment}

\hypersetup{
    pdfnewwindow=true,      
    colorlinks=true,       
    linkcolor=black,          
    citecolor=blue,        
    filecolor=blue,      
    urlcolor=blue           
}

\def\bea{\begin{eqnarray}}
\def\eea{\end{eqnarray}}

\begin{document}
\title{Emerging Internal Symmetries from Effective Spacetimes}
\author{Manfred Lindner and Sebastian Ohmer}
\address{Max-Planck-Institut f\"{u}r Kernphysik, Saupfercheckweg 1, D-69117 Heidelberg, Germany}
\begin{abstract}
Can global internal and spacetime symmetries be connected without supersymmetry? To answer this question, we investigate Minkowski spacetimes with $d$ space-like extra dimensions and point out under which general conditions external symmetries induce internal symmetries in the effective $4$-dimensional theories. We further discuss in this context how internal degrees of freedom and spacetime symmetries can mix without supersymmetry in agreement with the Coleman-Mandula theorem. We present some specific examples which rely on a direct product structure of spacetime such that orthogonal extra dimensions can have symmetries which mix with global internal symmetries. This mechanism opens up new opportunities to understand global symmetries in particle physics.
\end{abstract}
\maketitle
\section{Introduction}
The nature of spacetime is still a great mystery in fundamental physics and it might be a truly fundamental quantity or it could be an emergent concept. An appealing and most minimalistic approach would be if spacetime and propagating degrees of freedom would have a common origin on equal footing. In such a scenario, spacetime is thus an emergent quantity and there seems to be no reason for it to be restricted to a $4$-dimensional Poincar\'e symmetry apart from low energy phenomenology. The only exception are additional time-like dimensions which typically lead to inconsistencies when requiring causality~\cite{Time1, Time2}, while there is no consistency problem with additional space-like dimensions. Additional space-like dimensions have therefore been widely studied. 

If spacetime and particles consist of the same building blocks, then a fundamental connection of these low energy quantities should exist at high energies. Early attempts in this direction have led to the Coleman-Mandula no-go theorem~\cite{CMTheorem}. The no-go theorem shows under general assumptions that a symmetry group accounting for $4$-dimensional Minkowski spacetime and internal symmetries has to factor into the direct product of spacetime and internal symmetries. This implies that spacetime and particle symmetries cannot mix in relativistic interacting theories.

One way to circumvent the no-go theorem is to study graded symmetry algebras which introduce fermionic symmetry generators and are known as supersymmetries~\cite{SUSY}. The possibility to mix spacetime and internal symmetries in a relativistic theory is a strong theoretical argument for supersymmetry and supersymmetric extensions of the Standard Model of particle physics are therefore widely studied. However, there is no experimental evidence for supersymmetry, see e.g.~\cite{susyLHCpro,susyLHC,susyDM}, and it is a finely question to ask: Are there alternative ways to circumvent the Coleman-Mandula theorem?

The answer to this question is: Yes. We therefore relax the assumption that spacetime is described by the $4$-dimensional Poincar\'e symmetry. We then investigate new alternative scenarios to mix global spacetime and internal symmetries. Next, we review the Coleman-Mandula theorem to understand how to circumvent the theorem with extra space dimensions. In section \ref{trans}, we discuss translational invariant extra dimensions and show how momentum conservation can be interpreted as new internal symmetry. We then go further in section \ref{rot} and consider extra dimensions described by rotational invariant spacetimes which lead to ``hidden'' spins. Finally, we investigate how rotational and internal symmetries can mix if the rotational symmetry group is compact in section \ref{nonrel}. Such scenarios can for example lead to an explanation of the three Standard Model families. We conclude and give an outlook for further investigations in section \ref{con}.

\section{Coleman-Mandula no-go theorem} 
\label{CM}

The Coleman-Mandula theorem~\cite{CMTheorem, Horwitz, Weinberg} states, if $G$ is a connected symmetry group of the S-matrix and
\begin{compactenum}[(i)]
\item $G$ has a subgroup which is locally isomorphic to the Poincar\'e group,
\item all physical particles have positive definite mass and there exists only a finite number of particles below an energy threshold $E_\text{max}$,
\item the S-matrix is an analytic function of the Mandelstam variables $s$ and $t$,
\item the S-matrix is non-trivial at almost all energies,
\item generators of $G$ are representable as integrals in momentum space,
\end{compactenum}
then $G$ is locally isomorphic to the Poincar\'e group times an internal symmetry group.

It is important to develop a physical intuition for the Coleman-Mandula theorem~\cite{Witten, susy2001}. A physical scattering amplitude has to respect all symmetries of the theory and thus the number of independent variables describing the scattering process is reduced. Requiring that the theory respects the laws of special relativity implies that the scattering amplitude is a Lorentz scalar. Moreover, for a scattering process to be physical the initial and final $4$-momenta have to be on the mass-shell. We further demand that the scattering process respects energy-momentum conservation. Taking into account all these kinematic restrictions for a 2$\to$2 scattering process only leaves the famous Mandelstam variables $s$ and $t$ as free parameters in $d+1$ dimension with $d > 1$. If we would demand that the scattering process respects an additional conserved charge which is a function of the momenta, then only discrete scattering angles would be allowed. This is however in conflict with the assumptions since scattering should be non-trivial for most energies. We therefore can conclude that further restrictions on the scattering amplitudes should be independent of the $4$-momenta of the particles. We would call such a symmetry an internal symmetry since its generator would commute with the spacetime generators. This implies for the general symmetry structure of the S-matrix
\begin{align}
G \to \mathcal{P}(1,3) \otimes \text{``internal symmetries''}\,.
\end{align}
A more detailed mathematical treatment can be found in~\cite{CMTheorem, Horwitz, Weinberg} but the essence is that Lorentz invariance severely restricts the possible symmetries of the S-matrix.

However, if we extend the underlying $4$-dimensional Poincar\'e invariant spacetime by $d$ space-like dimensions, where we assume that the symmetry generators commute with the $4$-dimensional Poincar\'e group, the scattering process is allowed to respect conserved charges which depend on the momenta in the $d$ space-like dimensions without discretizing the $4$-dimensional Mandelstam variables. We will implement this in section \ref{trans}.

Until now we only considered scattering of scalar degrees of freedom which transform trivially with respect to Lorentz transformations. However, introducing particles with spin, we introduce degrees of freedom which transform non-trivially with respect to Lorentz transformations. We can now ask if there is a conserved charge of a scattering process which depends on the spin of the particles. Such a conserved charge would belong to symmetry transformations which relate particles in different representations of the Lorentz group. Such symmetries were considered in the context of relativistic $SU(6)$ theories~\cite{rSU6A,rSU6B,rSU6C} and it was soon pointed out that such theories have unfavorable features such as an infinite number of particles~\cite{dSU6A,dSU6B,dSU6C}. The root of these problems is the structure of the Lorentz group. The semi-simple part of the Lorentz group, $SO(1,3)$, is non-compact due to the unbounded Lorentz boosts and thus does not have non-trivial unitary finite-dimensional representations. When comparing to non-relativistic spacetimes which transform according to the Galilean symmetry group, we do not observe such inconsistencies~\cite{Wigner}. This is due to the compactness of the semi-simple part of the Galilean group, $SO(3)$.

Utilizing this property, we will introduce additional space-like dimensions which transform rotationally according to a compact symmetry group in section \ref{nonrel}. We then illustrate how global spacetime and internal symmetries can mix. This can then give rise to new symmetries which may be the origin of the family and flavor structure of the Standard Model.

\section{Translational Symmetries}
\label{trans}

First, we consider the simple and well known example of a $D$-dimensional theory, $D = 4 +d$, with $d$ extra dimensions where spacetime is described by 
\begin{align}
\mathcal{M}_4 \times \Sigma_d \,,
\end{align}
with $\mathcal{M}_4$ the $4$-dimensional Minkowski spacetime and $\Sigma_d$ the additional $d$-dimensional space. The spacetime coordinates can thus be written as $z^A = (x^\mu, y^a)$ with $\mu = (0,1,2,3)$ and $a=(4,\dots,D-1)$. The spacetime symmetry group factors as
\begin{align}
\mathcal{P}(1,3) \otimes G_d \,,
\end{align}
where $\mathcal{P}(1,3)$ is the $4$-dimensional Poincar\'e group and $G_d$ is the symmetry group of $\Sigma_d$. We further assume that the space described by $G_d$ is translational invariant such that the $(4+d)$-dimensional momentum
\begin{align}
P^A = \int \mathrm{d}^3 x\, \mathrm{d}^d y\, T^{0 A}\,,
\end{align}
with $A = (0,1,\dots, D-1)$ and energy-momentum tensor $T^{AB}$ is conserved, $\partial_0 P^A = 0$. We also assume that
\begin{align}
m^2 = P^\dagger_A P^A \quad \text{with} \quad A = (0,1,\dots, D-1) \,,
\end{align}
commutes with all group generators and that $m^2$ is a constant for all irreducible representations. The particles momenta in the extra dimension thus contribute to the energy-momentum relation
\begin{align}
E^2 = m^2 + |\vec{p}|^2 + \left( p_4^2 + \dots + p_{D-1}^2 \right) \,,
\end{align}
although the generators $P^a$ with $a \in (4, \dots , D-1)$ commute with all generators of the Poincar\'e group $\mathcal{P}(1,3)$ and would thus naively account for internal symmetries.

The assumed spacetime structure gives rise to additional conserved charges connected to the particle momenta in the extra dimensions $P^a$ with $a \in (4, \dots, D-1)$. Scattering processes will then have to respect additional conservation laws. The schematic scattering process
\begin{align}
(\vec{p}_A, p_D) + (\vec{p}_B,0) \rightarrow (\vec{p}_A, 0) + (\vec{p}_B,0) \,,
\end{align}
would for example be forbidden. Note that the new conserved charges will not discretize the $4$-dimensional scattering process. Moreover, from a $4$-dimensional point of view the scattering process respects additional internal symmetries. We cannot distinguish from the scattering process if the additional symmetry is due to an enriched spacetime structure, or due to additional internal symmetries. This is not a contradiction to the Coleman-Mandula theorem. The factorization of the general symmetry group of the S-matrix $G$ can also include additional spacetime symmetries
\begin{align}
G \to \mathcal{P}(1,3) \otimes G_d \otimes \text{``internal symmetries''}\,.
\end{align}

A prime example for such a symmetry are Kaluza-Klein numbers in theories with universal extra dimensions~\cite{UED, KKrad} which can stabilize dark matter~\cite{KKDM, Cacciapaglia}. The Kaluza-Klein number is no longer a continuous observable such as the momenta discussed above. The extra dimensions have to be compactified in phenomenologically viable models which breaks the translational invariance of the extra dimensions. After compactification, the translational invariance of the $d$ space-like dimensions is not conserved globally, but only locally in space. The momenta in the extra dimensions can only take discrete values. Quantum corrections to the particles mass will depend on the momentum in the extra dimensions due to non-local loop contributions~\cite{KKrad}. Note that for orbifold compactifications translational invariance is further broken by the orbifold fixpoints. Scattering processes then have to conserve Kaluza-Klein parity~\cite{KKrad}.

\section{Rotational Symmetries}
\label{rot}

The $4$-dimensional spacetime we observe is not only translational invariant but also rotational invariant. It seems therefore natural to consider rotational symmetries in extra dimensions. Let us thus move to the scenario where $G_d$ further incorporates rotational symmetries. The simplest scenario has two extra space-like dimensions, $d=2$. We hence assume that the additional space dimensions are given by $\Sigma_2 \cong \mathbb{R}^2$ and the spacetime symmetry which describes $\Sigma_2$ is given by $G_2 \cong \mathbb{R}^2 \rtimes SO(2)$. $G_2$ does now generate translational and rotational symmetries. The full spacetime structure is thus given by
\begin{align}
\mathcal{M}_4 \times \mathbb{R}^2 \,,
\end{align}
with spacetime symmetry
\begin{align}
\mathcal{P}(1,3) \otimes \left( \mathbb{R}^2 \rtimes SO(2) \right) \,.
\end{align}
Again, we find two additional conserved momenta
\begin{align}
\partial_0 P^4 = 0 \quad \text{and} \quad \partial_0 P^5 = 0 \,.
\end{align}
Furthermore, the angular momentum $L^{45} = y^4 p^5 - y^5 p^4$ in the plane $\Sigma_2$ is also conserved
\begin{align}
\partial_0 L^{45} = 0 \,.
\end{align}
Since we focus on particle interactions and their symmetries we can always choose a reference frame where the initial angular momentum is zero.\footnote{But note that the additional overall angular momentum conservation could have interesting implications for finite temperature dynamics in early universe cosmology.}

We also have to take into account that particle wave functions do not have to transform trivially under space rotations. To illustrate this point, we consider the rotation $R^{45}(\theta)$ in the plane $\Sigma_2$ given by
\begin{align}
\begin{pmatrix} {y'}^4 \\ {y'}^5 \end{pmatrix} = \begin{pmatrix} \text{cos}(\theta) && -\text{sin}(\theta) \\ \text{sin}(\theta) && \text{cos}(\theta) \end{pmatrix} \begin{pmatrix} {y}^4 \\ {y}^5 \end{pmatrix}\,.
\end{align}
The wave function of a particle in a non-trivial representation with respect to $G_2$ then transforms under the rotation $R^{45}(\theta)$ as
\begin{align}
\Psi(x^\mu, y^4, y^5) \rightarrow e^{-i \theta s_h}\,\Psi(x^\mu, y^4, y^5) \,. 
\end{align}
We can thus introduce the ``hidden'' spin $s_h$ in the extra plane which, for a two dimensional space, can take values $s_h \in \mathbb{R}$. Particles therefore behave as anyons in the additional space dimensions. However, from a $4$-dimensional perspective, this transformation corresponds to a global $U(1)$ symmetry. In other words: The $4$-dimensional $U(1)$ charge of the particle can be identified with the ``hidden'' spin $s_h$. The $U(1)$ symmetry related to the non-trivial transformation of the particle wave function under rotations in $\Sigma_2$ is indistinguishable from global internal $U(1)$ symmetries in $4$-dimensions.

Such a rotational symmetry on the extra dimensional space $\Sigma_d$ which induces a symmetry in the effective $4$-dimensional theory is known from the compactification on the chiral square~\cite{Dobrescu, Proton6D}. To have a realistic theory which includes chiral fermions the compactification further requires orbifolding. The folding boundary conditions break the continuous rotational symmetry $U(1)$ to a discrete rotational symmetry $Z_8$.

\section{Mixed Symmetries}
\label{nonrel}

So far we have only considered how spacetime symmetries induce internal symmetries in the effective $4$-dimensional theory. However, we are mostly interested in scenarios where global internal and spacetime symmetries mix in agreement with the Coleman-Mandula theorem. We therefore have to demonstrate how global internal and spacetime symmetries can be combined in a single global symmetry. Following the example of non-relativistic theories where spacetime is characterized by the Galilean group~\cite{Wigner}, we require that the space rotations in $\Sigma_d$ are described by a compact subgroup of $G_d$. We can then construct mixed global symmetries where particles with different ``hidden'' spin $s_h$ are in the same multiplet. To illustrate this new aspect and make a connection to the previous example, we assume that at high energies spacetime is given by
\begin{align}
\mathcal{M}_4 \times \mathbb{R}^3\,.
\end{align}

In this section, we assume that spacetime is not an ordinary manifold. Moreover, we assume that spacetime arises effectively from a more fundamental theory and that there is an effective structure, such as a condensate, which allows locally to distinguish $\mathcal{M}_4$ and $\mathbb{R}^3$.\footnote{We thank Arthur Hebecker for clarification.} Moreover, we explicitly assume that $\mathcal{M}_4 \times \mathbb{R}^3$ does not originate from $\mathcal{M}_7$.\footnote{Hence, the theory is intrinsically not higher dimensional Poincar\'e invariant.} We then propose to interpret elementary particles as irreducible representations of the global spacetime symmetry of $\mathcal{M}_4 \times \Sigma_d$. The global spacetime symmetry of $\mathcal{M}_4 \times \mathbb{R}^3$ would thus be $\mathcal{P}(1,3) \otimes G_3$ with $G_3 \cong \mathbb{R}^3 \rtimes SU(2)$. However, to illustrate the mixing of global spacetime symmetries and internal symmetries, we further assume that the spacetime symmetry $\mathcal{P}(1,3) \otimes G_3$ is now given by $G_3 \cong \mathbb{R}^3 \rtimes SU(3)$. The global $SU(3)$ symmetry mixes an internal global $U(1)_I$ symmetry and the rotational spacetime symmetry described by the compact group $SU(2)$
\begin{align}
SU(3)\supset U(1)_I \otimes SU(2)\,.
\end{align}

As a toy model, we consider the fermionic field $\Psi$ which transforms as a spin $\frac{1}{2}$-representation of $\mathcal{P}(1,3)$ and a ``hidden'' spin $\frac{1}{2}$-representation of $G_3$.\footnote{In principle, all possible combinations of irreducible representations of $\mathcal{P}(1,3)$ and $G_3$ are allowed.} The field $\Psi$ is thus given by
\begin{align}
\Psi^{F\,f} \hspace{-1pt}(x^\mu, y^i) = \psi^F \hspace{-1pt}(x^\mu) \psi^f \hspace{-1pt}(y^i) \quad \text{with} \quad i \in (1, 2, 3) \,,
\end{align}
with the $4$-dimensional spinor index $F \in (0, 1, 2, 3)$. The spinor index $f$ would range in $f \in (1, 2)$ if $\psi^f \hspace{-1pt} (y^i)$ would transform according to the fundamental representation of $SU(2)$. However, since the spacetime symmetry is assumed to be the mixed symmetry $SU(3)$ the spinor index $f$ ranges in $f \in (1, 2, 3)$ for $\psi^f \hspace{-1pt} (y^i)$ in the fundamental representation of $SU(3)$. The action of a free fermion is thus given by
\begin{align}
S = \int \mathrm{d}^4 x\, \mathrm{d}^3 y \Bigg(& \bar{\Psi}_{F\,f} \left( i \left(\gamma^\mu\right)^F_G \partial_\mu - \frac{\delta^F_G}{2 M} \partial_i \partial^i \right) \Psi^{G\,f} \nonumber \\ 
&- m\, \bar{\Psi}_{F\,f} \Psi^{F \, f} \Bigg) \,,
\end{align}
where $\bar{\Psi}_{F\,f}(x^\mu, y^i) = \bar{\psi}_F(x^\mu) \psi^\dagger_f(y^i)$. The action can be further simplified
\begin{align}
S =&\, N \int \mathrm{d}^4 x  \Big(\bar{\psi}_{F}(x^\mu) \left( i \left(\gamma^\mu\right)^F_G \partial_\mu - m\, \delta^F_G \right) \psi^{G} \hspace{-1pt} (x^\mu) \Big)\nonumber \\ 
+ \int & \mathrm{d}^4 x\, \bar{\psi}_F(x^\mu)\psi^F \hspace{-1pt} (x^\mu) \int \mathrm{d}^3 y\,  \psi_f^\dagger(y^i) \left(-\frac{1}{2M}\partial_i \partial^i \right)\psi^f \hspace{-1pt} (y^i)  \,,
\end{align}
with normalization constant
\begin{align}
N = \left(\int \mathrm{d}^3 y\, \psi_f^\dagger(y^i) \psi^f \hspace{-1pt} (y^i)\right) \,.
\end{align} 
Note that the ``non-relativistic kinetic term'', $(1/2M) \partial_i \partial^i$, for the ``static'' field $\psi^f \hspace{-1pt} (y^i)$ is due to the ``non-relativistic'' spacetime symmetry $G_3$.

The global ``non-relativistic'' spacetime symmetry $SU(3)$ acts on $\Psi^{F\,f} \hspace{-1pt} (x^\mu, y^i)$ as
\begin{align*}
\Psi^{F\,f} \hspace{-1pt} (x^\mu, y^i) \rightarrow (e^{-i \alpha_N \lambda^N})^f_g \Psi^{F\,g} \hspace{-1pt} (x^\mu, y^i) \,,
\end{align*}
with $N \in (1, \dots, 8)$, $\alpha_N$ finite group parameters and $\lambda^N$ the Gell-Mann matrices.

The fact that the $SU(3)$ symmetry mixes global internal and spacetime symmetries becomes evident upon compactification of one extra dimension onto a circle. The spacetime structure breaks down to 
\begin{align}
\mathcal{M}_4 \times \mathbb{R}^3 \to \mathcal{M}_4 \times \mathbb{R}^2 \times S^1 \,,
\end{align}
and thus the remaining spacetime symmetries are $\mathcal{P}(1,3) \otimes G_2$ with $G_2 \cong \mathbb{R}^2 \rtimes U(1)_S$. The former global $SU(3)$ symmetry is now broken to a global internal $U(1)_I$ symmetry and an $1$-dimensional spacetime rotational symmetry $U(1)_S$
\begin{align}
SU(3) \to U(1)_I \otimes U(1)_S \,.
\end{align}

The ``static field'' $\psi^f \hspace{-1pt} (y^i)$ can now be expanded as
\begin{align}
\psi^f (y^j, y^3) = \frac{1}{\sqrt{2 \pi R}} \sum_l {\psi^{(l)}}^f \hspace{-1pt} (y^j)\, e^{i \frac{l}{R}y^3} \,,
\end{align}
with $j \in (1,2)$ and $R$ the compactification radius of $S^1$. The second term of the action thus simplifies to
\begin{align}
S \supset & \int \mathrm{d}^4 x\, \bar{\psi}_F(x^\mu)\psi^F \hspace{-1pt} (x^\mu) \int \mathrm{d}^2 y\, \sum_l {\psi_f^{(l)}}^\dagger \hspace{-1pt} (y^j) \nonumber \\
& \times \left(-\frac{1}{2M}\partial_j \partial^j - \frac{l^2}{2 M R^2} \right){\psi^{(l)}}^f \hspace{-1pt} (y^j) \,.
\end{align}
We can thus define
\begin{align}
M_1 &= \int \mathrm{d}^2 y\, \sum_l {\psi_1^{(l)}}^\dagger \hspace{-1pt} (y^j)\left(-\frac{1}{2M}\partial_j \partial^j - \frac{l^2}{2 M R^2} \right){\psi^{(l)}}^1 \hspace{-1pt} (y^j) \,,\nonumber \\
M_2 &= \int \mathrm{d}^2 y\, \sum_l {\psi_2^{(l)}}^\dagger \hspace{-1pt} (y^j)\left(-\frac{1}{2M}\partial_j \partial^j - \frac{l^2}{2 M R^2} \right){\psi^{(l)}}^2 \hspace{-1pt} (y^j) \,,\nonumber \\
M_3 &= \int \mathrm{d}^2 y\, \sum_l {\psi_3^{(l)}}^\dagger \hspace{-1pt} (y^j)\left(-\frac{1}{2M}\partial_j \partial^j - \frac{l^2}{2 M R^2} \right){\psi^{(l)}}^3 \hspace{-1pt} (y^j)\,,
\end{align}
and further require
\begin{align}
m_1 &= m - M_1\,, \nonumber \\
m_2 &= m - M_2 \,,\nonumber \\
m_3 &= m - M_3 \,,
\end{align}
where $m_1$, $m_2$ and $m_3$ are the experimentally measured masses of the three Standard Model fermion generations. The different Standard Model fermion masses are therefore due to the different field configurations of ${\psi^{(l)}}^1 \hspace{-1pt} (y^j)$, ${\psi^{(l)}}^2 \hspace{-1pt} (y^j)$ and ${\psi^{(l)}}^3 \hspace{-1pt} (y^j)$.

The fundamental representation of the global $SU(3)$ breaks down such that
\begin{align}
\textbf{3} \to (1,(s_h = \frac{1}{2}))+ (1,(s_h = -\frac{1}{2})) + (-2 ,(s_h = 0))\,, \nonumber
\end{align}
which illustrates that states of different ``hidden'' spins were mixed in the $SU(3)$ multiplet. The individual  components of the ``static'' field ${\psi^{(l)}}^f \hspace{-1pt} (y^j)$ thus transform with respect to $U(1)_S$ as
\begin{align}
{\psi^{(l)}}^1 \hspace{-1pt} (y^j) &\rightarrow e^{-i \frac{\alpha_3}{2}} {\psi^{(l)}}^1 \hspace{-1pt} (y^j) \,, \nonumber \\
{\psi^{(l)}}^2 \hspace{-1pt} (y^j) &\rightarrow e^{i \frac{\alpha_3}{2}} {\psi^{(l)}}^2 \hspace{-1pt} (y^j) \,, \nonumber \\
{\psi^{(l)}}^3 \hspace{-1pt} (y^j) &\rightarrow {\psi^{(l)}}^3 \hspace{-1pt} (y^j) \,,
\end{align}
and with respect to $U(1)_I$ as 
\begin{align}
{\psi^{(l)}}^1 \hspace{-1pt} (y^j) &\rightarrow e^{-i \frac{\alpha_8}{2 \sqrt{3}}} {\psi^{(l)}}^1 \hspace{-1pt} (y^j) \,, \nonumber \\
{\psi^{(l)}}^2 \hspace{-1pt} (y^j) &\rightarrow e^{-i \frac{\alpha_8}{2 \sqrt{3}}} {\psi^{(l)}}^2 \hspace{-1pt} (y^j) \,, \nonumber \\
{\psi^{(l)}}^3 \hspace{-1pt} (y^j) &\rightarrow e^{i \frac{\alpha_8}{\sqrt{3}}} {\psi^{(l)}}^3 \hspace{-1pt} (y^j) \,.
\end{align}

We interpret the discrete ``hidden'' spin index $f \in (1, 2, 3)$ as generation index. The different mass contributions of the ``static'' fields ${\psi^{(l)}}^f(y^i)$ to $\psi^F(x^\mu)$ can thus be interpreted as the appearance of three generations with different masses in the effective $4$-dimensional theory. This example demonstrates how ``hidden'' spins might explain the appearance of three copies of fermions which could be identified with the three generations in the Standard Model of particle physics. By assuming spacetime is described by $\mathcal{M}_4 \times \mathbb{R}^3$ with a global mixed $SU(3)$ symmetry where the Standard Model fermions transform in the fundamental representation, we automatically find three copies of Standard Model fermions at low energies. It is important to note that the appearance of three generations is a consequence of the transformation property of the Standard Model fermions with respect to a mixed symmetry. Moreover, such ``hidden'' spins could also lead to viable explanations for the flavor~\cite{Flavor1, Flavor2, Flavor3, Flavor4, Flavor5} and family~\cite{Family1, Family2, Family3, Family4, Family5} structure of the Standard Model.

A different mechanism which relates flavor symmetries to spacetime symmetries was discussed in~\cite{Altarelli, Adulpravitchai}. The discrete flavor symmetries arise as a remnant of 6-dimensional Poincar\'e symmetry. Upon compactification via orbifolding and identifying the $4$-dimensional branes at the orbifold fixed points with representations of a discrete symmetry group such as $A_4$, a connection to discrete flavor symmetries is established. The number of fermionic generations in the Standard Model was derived from anomaly cancellation in a $6$-dimensional Lorentz invariant theory in~\citep{Anomaly6D}.

Note that this scenario is again no contradiction to the Coleman-Mandula theorem. We are exploiting the fact that the additional symmetries of the S-matrix which appear in the direct product with the Poincar\'e symmetry can be a mixture of additional non-relativistic spacetime symmetries and common internal symmetries
\begin{align}
G \to \mathcal{P}(1,3) \otimes \text{``mixed $G_d$ and internal symmetries''}\,.
\end{align}
It is important to stress again that the spacetime symmetries $G_d$ can only mix with global internal symmetries if $G_d$ contains a compact subgroup which gives rise to non-trivial unitary finite-dimensional representations.

\section{Conclusions \& Outlook}
\label{con}

In this paper, we discuss how to connect in principle global spacetime and internal symmetries without supersymmetry. The construction of phenomenologically viable models based on the toy models presented is left for future work. Phenomenologically viable models are more complex and  evolved since interactions have to be included. However, we have pointed out how spacetime extensions can give rise to internal symmetries and further how it is possible to mix spacetime and internal symmetries in rotational compact extra dimensions without supersymmetry. All presented extensions are in full agreement with the Coleman-Mandula theorem. These mechanisms could also be used to explain the stability of dark matter, the flavor structure of the Standard Model or give a physical reason for the three fermionic generations in the Standard Model as our example in section \ref{nonrel} illustrates.

There exist very good reasons which make a connection between internal symmetries and spacetime symmetries attractive. The discovery of supersymmetry would establish such a connection and it would show that four spacetime dimensions are sufficient. One could even argue that this could be considered as a strong hint towards the fundamental nature of a $4$-dimensional Minkowski spacetime. 

In this paper, we point out that internal symmetries and spacetime symmetries could still be connected even if supersymmetry is not discovered. For that, one has to rethink the structure of spacetime such that connections between global internal and spacetime symmetries emerge.

To illustrate that point, we used spacetimes for simplicity which are the direct product of $4$-dimensional Minkowski spacetime and $d$ extra space dimensions. We discussed a class of theories which circumvent the Coleman-Mandula theorem by arriving at a $4$-dimensional Minkowski spacetime from higher dimensions by compactification. This led to interesting possibilities how, for example, hidden spins could ultimately be related to fermion generations. 
We would like to emphasize that this work should be viewed as a step towards more general cases. Future work is also necessary to understand field theories which can lead to spacetimes which are a direct product of Minkowski spacetime and $d$ extra space dimensions without relying on higher dimensional Poincar\'e invariance.

One possible scenario where no extra dimensions are required are theories where the Poincar\'e symmetry emerges only effectively at low energies, while it is absent at the fundamental level. An example of such a theory is Horava-Lifshitz gravity which is intrinsically non-relativistic at high energies~\cite{Horava}.  
Another route could be to add extra dimensions where more complicated dynamical mechanisms lead to the emergence of $4$-dimensional Minkowski spacetime. 
Altogether, we conclude that even without supersymmetry interesting connections between internal symmetries and spacetime symmetries could exist. 

\section{Acknowledgment} 
We thank Gia Dvali, Florian Goertz, Arthur Hebecker and Hermann Nicolai for very useful discussions and comments on the manuscript.



\bibliography{ref_CM}{}
\bibliographystyle{utcaps_mod}

\end{document}